\documentclass[lettersize,journal]{IEEEtran} 

\ifCLASSOPTIONcompsoc
  \usepackage[nocompress]{cite}
\else
  \usepackage{cite}
\fi

\usepackage[ruled,vlined,lined,commentsnumbered,linesnumbered]{algorithm2e}

\usepackage{enumitem}
\usepackage{booktabs}
\usepackage{moreverb}
\usepackage{fontenc}
\usepackage{amsmath}

\usepackage{amssymb}
\usepackage{fancybox}
\usepackage{color}
\usepackage{colortbl}
\usepackage{array}
\usepackage{multirow}
\usepackage{multicol}
\usepackage{listings}
\usepackage{makecell}
\usepackage{graphicx}
\usepackage{setspace}
\usepackage{soul}
\usepackage[breaklinks]{hyperref}
\usepackage{balance}
\usepackage{csvsimple}
\usepackage{longtable}
\usepackage[skip=0pt,font=small,labelfont=bf]{caption}
\usepackage{url}
\usepackage{lscape}
\usepackage{rotating}
\usepackage{tikz}
\usepackage[normalem]{ulem}
\usepackage{subcaption}
\usepackage{wrapfig}
\usepackage[most]{tcolorbox}
\newcounter{myboxcounter}

\usepackage{marvosym}
\usepackage{pifont}

\newcolumntype{L}[1]{>{\raggedright\let\newline\\\arraybackslash\hspace{0pt}}m{#1}}
\newcolumntype{C}[1]{>{\centering\let\newline\\\arraybackslash\hspace{0pt}}m{#1}}
\newcolumntype{R}[1]{>{\raggedleft\let\newline\\\arraybackslash\hspace{0pt}}m{#1}}

\definecolor{codegreen}{rgb}{0,0.6,0}
\definecolor{codegray}{rgb}{0.5,0.5,0.5}
\definecolor{codepurple}{rgb}{0.58,0,0.82}
\definecolor{backcolour}{rgb}{0.95,0.95,0.92}
\definecolor{lightgreen}{HTML}{99d8c9}
\definecolor{lightgreen2}{HTML}{CCD2CC}
\definecolor{lightred}{HTML}{F4C7B0} 

\lstdefinestyle{mystyle}{
    commentstyle=\color{codegreen},
    keywordstyle=\color{magenta},
    numberstyle=\tiny\color{black},
    stringstyle=\color{codepurple},
    basicstyle=\footnotesize,
    breakatwhitespace=false,
    breaklines=true,
    captionpos=b,
    keepspaces=true,
    showspaces=false,
    showstringspaces=false,
    showtabs=false,
    tabsize=2
}

\definecolor{darkpastelred}{rgb}{0.76, 0.23, 0.13}
\definecolor{ao(english)}{rgb}{0.0, 0.5, 0.0}

\definecolor{darkpastelred}{rgb}{0.76, 0.23, 0.13}
\definecolor{ao(english)}{rgb}{0.0, 0.5, 0.0}
\lstdefinelanguage{diff}{
	morecomment=[f][\color{blue}]{@@},     
	morecomment=[f][\color{red}]-,         
	morecomment=[f][\color{codegreen}]+,       
	morecomment=[f][\color{red}]{---}, 
	morecomment=[f][\color{codegreen}]{+++},
}

\definecolor{yellow}{RGB}{255,255,153}
\definecolor{grey}{RGB}{224,224,224}

\newcommand{\prompt}[3]{
\refstepcounter{myboxcounter}
\begin{tcolorbox}[colback=gray!10, colframe=black!80,
width=\linewidth, arc=2mm, auto outer arc, title={{\small #1}}, label={prompt:#2}, center, left=2mm,right=2mm]
{
\begingroup
\small
\vspace{-2mm}
\linespread{0.7}\selectfont
#3
\vspace{-2mm}
\endgroup
}
\end{tcolorbox}
}

\newboolean{showcomments}
\setboolean{showcomments}{true}
\ifthenelse{\boolean{showcomments}}
 { \newcommand{\mynote}[2]{
      \fbox{\bfseries\sffamily\scriptsize#1}
        {\small$\blacktriangleright$\textsf{\emph{#2}}$\blacktriangleleft$}}}
        { \newcommand{\mynote}[2]{}}

\definecolor{DarkOrange}{rgb}{0.8,0.3,0.0}
\definecolor{DarkCyan}{rgb}{0.0, 0.55, 0.55}

\newcolumntype{?}{!{\vrule width 1pt}}

\definecolor{grey}{rgb}{0.9,0.9,0.9}
\definecolor{lightgrey}{HTML}{f0f0f0}
\definecolor{mygreen}{HTML}{02818a}
\definecolor{mygray}{HTML}{666666}

\newcommand{\toolname}{\texttt{LEADER}\xspace}
\newcommand*{\eg}{e.g., }

\newcommand*{\ie}{i.e., }

\newcommand*{\etal}{et al. }

\newcommand{\update}[1]{\textcolor{blue}{#1}}

\ifCLASSINFOpdf
\else
\fi
\begin{document}
%
\title{Large Language Models-Aided Program Debloating}
\pagenumbering{arabic}

\author{
    \IEEEauthorblockN{
        Bo Lin,
        Shangwen Wang,
        Yihao Qin,
        Liqian Chen,
        Xiaoguang Mao
    }
    
    \IEEEauthorblockA{National University of Defense Technology, Changsha, China,\\
    \{linbo19, wangshangwen13, yihaoqin, lqchen, xgmao\}@nudt.edu.cn
    }
     \IEEEcompsocitemizethanks{
     \IEEEcompsocthanksitem Shangwen Wang is the corresponding author.
     }
}
\maketitle

\begin{abstract} 
As software grows in complexity to accommodate diverse features and platforms, software bloating has emerged as a significant challenge, adversely affecting performance and security. 
However, existing approaches inadequately address the dual objectives of debloating: maintaining functionality by preserving essential features and enhancing security by reducing security issues. 
Specifically, current software debloating techniques often rely on input-based analysis, using user inputs as proxies for the specifications of desired features. However, these approaches frequently overfit provided inputs, leading to functionality loss and potential security vulnerabilities.
To address these limitations, we propose \toolname, a program debloating framework enhanced by Large Language Models (LLMs), which leverages their semantic understanding, generative capabilities, and decision-making strengths. \toolname mainly consists of two modules: (1) a documentation-guided test augmentation module designed to preserve functionality, which leverages LLMs to comprehend program documentation and generates sufficient tests to cover the desired features comprehensively, and (2) a multi-advisor-aided program debloating module that employs a neuro-symbolic pipeline to ensure that the security of the software can be perceived during debloating. This module combines debloating and security advisors for analysis and employs an LLM as a decision-maker to eliminate undesired code securely.
Extensive evaluations on widely used benchmarks demonstrate the efficacy of \toolname. It achieves a 95.5\% test case pass rate and reduces program size by 42.5\%. Notably, it reduces the introduction of vulnerabilities during debloating by 79.1\% and decreases pre-existing vulnerabilities by 16.5\% more than CovA. These results demonstrate that \toolname surpasses the state-of-the-art tool CovA in functionality and security. These results underscore the potential of \toolname to set a new standard in program debloating by effectively balancing functionality and security.
\end{abstract}


\section{Introduction}
\label{sec:intro}
\IEEEPARstart{A}{s} software evolves to encompass a broader range of features and platforms, it naturally becomes more intricate and extensive. However, it is common for end-users to interact with only a limited set of these features~\cite{hibbs2009art}.
Consequently, software bloating becomes increasingly prevalent, which poses significant challenges to software performance and security as it not only hampers software efficiency~\cite{xu2010software,xu2009go,bhattacharya2011interplay} but also introduces unnecessary attack vectors~\cite{azad2019less,shacham2007geometry}.
To combat this, various debloating techniques have been developed to automatically identify and remove unused features, simplifying the software and enhancing efficiency~\cite{heo2018effective,qian2019razor,qian2020slimium,sharif2018trimmer,tang2021xdebloat,xin2022studying}.

During software debloating, the community-acknowledged core objectives are twofold~\cite{heo2018effective,xin2022studying}: firstly, maintaining {\bf functionality} by removing superfluous code while preserving the integrity of required features; and secondly, enhancing {\bf security} by minimizing the attack surface through the elimination of vulnerable code segments.
However, existing program debloating techniques fall short on achieving these two goals.
On one hand, existing approaches rely on different tests to depict the required features of the program, but {\bf it is difficult to create accurate and comprehensive specifications for desired software features} ({\bf Challenge 1}):
Similar to mainstream approaches in other domains, such as program repair~\cite{le2019automated}, most debloating studies (also known as input-based approaches~\cite{xin2022studying}) use tests constructed from user inputs as a proxy for desired feature specifications~\cite{heo2018effective, qian2019razor, qian2020slimium, xin2022studying}. 
These tests measure the consistency of program output before and after debloating.
However, since these tests are only an under-approximation of the desired features, this type of approach might result in a program that overfits the tests. In other words, the program may only function correctly when given the specific tests it was debloated on.
Our investigation of the debloated programs generated by the baselines examined in this paper for the {\tt Util} and {\tt Sir} benchmarks (see \S\ref{sec:benchmark}) reveals that 79\% of the required features were not correctly retained.
Another study by Qi \etal\cite{xin2022studying} observed a similar phenomenon that existing debloating techniques can fail on approximately 30\% to 40\% of unseen test cases.
A vivid example is that the state-of-the-art \textsc{Chisel}~\cite{heo2018effective} produces a debloated version of {\tt mkdir} based on the input ``{\tt -p foo/bar}'', but this debloated version fails to handle ``{\tt -p foo/bar/baz}'' due to the elimination of the loop for identifying directories~\cite{xin2022studying}.
This phenomenon arises from a deficiency in comprehending the software and its required features, as input-based approaches are tailored not to grasp these necessary features but to merely generate software capable of handling the features defined by provided tests.
On the other hand, existing approaches utilize coverage analysis and greedily remove code that is not covered, but {\bf they lack the insights into the security of the debloated software during the debloating process} ({\bf Challenge 2}):
Typically, an input-based software debloating approach utilizes code pruning to aggressively compress the program size and ensure the reduced program functions correctly for the provided tests \cite{heo2018effective,regehr2012test,sun2018perses}. Yet, these methods often overlook the possibility that eliminating certain security-related code fragments could introduce vulnerabilities.
For instance, a null pointer checker may be eliminated if the given tests do not invoke it, leading to possible null pointer dereferences. To gain a comprehensive understanding of this phenomenon, we conducted an exploratory experiment on six debloating tools across 25 programs (as discussed in \S\ref{sec:security}), which revealed that the tools introduced dozens of high-severity vulnerabilities and hundreds of moderately severe vulnerabilities for each program on average.

Recently, LLMs have demonstrated remarkable performance across various code-related tasks~\cite{xia2023keep,li2023assisting,lu2023llama}. Trained on billions of open-source code snippets, these models exhibit exceptional capabilities in multiple aspects. For instance, LLMs excel at understanding complex program documentation by effectively summarizing the content and identifying the purposes and constraints of functions~\cite{nam2024using,yuan2023evaluating}. Building on this understanding, LLMs demonstrate impressive generative abilities, producing efficient code and synthesizing comprehensive test cases~\cite{deng2023large,ryan2024code,chen2024chatunitest,xia2024fuzz4all}. For example, LLM-based fuzzing has achieved higher code coverage compared to traditional fuzzers~\cite{xia2024fuzz4all}. 
Beyond their understanding and generation capabilities, researchers have also found that systems like ``LLM-as-a-Judge'' perform comparably to human judgment across various tasks~\cite{zheng2023judging,huang2024empirical,chang2024survey}. These systems exhibit strong decision-making abilities, enabling them to evaluate information and produce well-reasoned decisions.

Motivated by the challenges of existing debloating approaches and the inspiring capacity of LLMs, we propose \toolname, an \underline{L}LM-aid\underline{E}d progr\underline{A}m \underline{DE}bloate\underline{R} that simultaneously leverages the strengths of LLMs on understanding, generation and decision-making to enhance the functionality and security of the debloated programs.
\toolname mainly consists of two modules, \ie documentation-guided test augmentation and multi-advisor-aided program debloating, that aim at addressing the aforementioned two challenges respectively.
\toolname begins by documentation-guided test augmentation, which first leverages the LLMs' capability on semantic understanding to comprehend the necessary features with the help of program documentation, and then generates sufficient tests to cover the desired features for the downstream debloating process. The newly created tests enhance coverage of the desired features, increasing the likelihood of their preservation during the debloating process.
Subsequently, we construct a multi-advisor-aided program debloating module, whose primary goal is to ensure that the security of the software can be perceived during debloating.
This neuro-symbolic module contains a debloating advisor and a security advisor, which are responsible for providing action suggestions and delineating potential consequences, respectively, during debloating. 
The debloating advisor identifies potentially unused code through coverage analysis, while the static analysis-based security advisor evaluates the recommendations from the debloating advisor to prevent the introduction of security issues.
After that, we use an LLM as the decision-maker to determine which code snippets can be eliminated and how to remove them securely.
To obtain the final results, we evaluate the debloated programs using the given tests. Programs that produce the same output as the original are considered output candidates.
Programs that fail are subjected to repeated debloating until they either pass the given tests or reach the maximum number of attempts.

We have conducted extensive evaluations for \toolname, illustrating its superiority over previous studies. On two widely-used benchmarks (\ie {\tt Util}~\cite{heo2018effective} and {\tt SIR}~\cite{sir2022bench}), \toolname achieves higher correctness compared to previous studies. Specifically, programs in {\tt Util} debloated by \toolname pass 96\% of test cases on average, while reducing program size by 54\%. In contrast, the state-of-the-art approach, CovA~\cite{xin2022studying}, achieves only 73\% correctness. Moreover, \toolname not only delivers a significant improvement in functionality but also enhances the security of debloated programs. Notably, \toolname reduces 16.5\% more vulnerabilities on average (from 354.8 to 396.4) and decreases vulnerabilities introduced by the debloating process by 79.1\% (from 364.9 to 76.1) compared to CovA.
Furthermore, our detailed analysis indicates that each proposed module plays a crucial role in the debloating process, illustrating the efficacy of our approach. In summary, our study makes the following contributions:

\begin{itemize}[leftmargin=*]
    \item {\bf New Dimension:} Our work highlights the limitations of existing debloating approaches in maintaining both functionality and security. We show that it is possible to achieve both objectives by leveraging LLM-aided test augmentation and a multi-advisor-aided debloating module, ensuring that unnecessary code can be securely eliminated without compromising the integrity of desired features.

    \item {\bf State-of-the-Art Tool}: We present a novel LLM-aided program debloating approach, \toolname, marking the first algorithm to integrate LLMs into the debloating process. Leveraging the capabilities of LLMs in understanding, generating, and decision-making, \toolname incorporates a documentation-guided test augmentation and a multi-advisor-aided debloating mechanism to ensure both functionality and security throughout the debloating process.

    \item {\bf Extensive Study}: We comprehensively evaluated \toolname on 25 programs, ranging from 382 lines to over 136,000 lines. The results demonstrate \toolname's impressive performance in debloating programs while maintaining security.
\end{itemize}

\section{Background}
\label{sec:bg}
\subsection{Program Debloating}
Modern software systems often include numerous features and support multiple platforms, leading to uncontrolled increases in sizes~\cite{qian2019razor}. Despite this complexity, individual users typically require only a small subset of these features~\cite{heo2018effective,xin2022studying}.
Software debloating is a promising technique that reduces the effort of manual debloating and automates the generation of debloated software. These approaches can be categorized into input-based and non-input-based techniques based on how required features are specified~\cite{xin2022studying}.

Input-based techniques necessitate a set of tests specifying which features should be retained~\cite{kalhauge2019binary,qian2019razor,heo2018effective,regehr2012test,xin2020program}. For example, \textsc{Chisel} employs delta debugging combined with a Markov decision process to remove unnecessary snippets from the code, while ensuring that the minimized program still satisfies specified properties. Similarly, \textsc{Razor} collects execution traces from tests and constructs a control flow graph (CFG) based on these traces. This graph is then expanded using heuristics to include potentially useful but initially uncovered paths, and a generator produces the debloated program based on the refined CFG.

Non-input-based techniques do not require tests to specify required features. Instead, they rely on other forms of specification. For example, \textsc{Carve}~\cite{brown2019carve} uses human annotations to specify code related to required features implicitly. Other specifications include human-developed domain sampling~\cite{xin2020subdomain}, configuration data~\cite{sharif2018trimmer}, and surveys~\cite{ali2011moms}.

This study focuses on input-based techniques, as non-input-based approaches rely heavily on external human input, such as developer annotations. Input-based techniques, in contrast, can leverage tests created during software development. However, these tests typically offer only an under-approximation of the required features, covering only a portion of the code needed in the software. Consequently, this limitation may compromise the functionality of the debloated program.
Existing input-based approaches mitigate functionality loss caused by inadequate specifications via strategies: inferring potentially required statements from historical data or augmenting tests. The first strategy utilizes techniques such as reinforcement learning~\cite{heo2018effective}, heuristics~\cite{qian2019razor}, and dependency analysis~\cite{xin2022studying} to infer uncovered but potentially necessary statements. However, insufficient use of historical knowledge results in over 30\% of previously unseen tests failing~\cite{xin2022studying}.
The second strategy, employed by tools like CovF~\cite{xin2022studying} uses fuzzing techniques to augment tests. However, fuzzing often produces tests that exhibit irregular patterns, exercising non-core logic, making it challenging to adequately cover the required features~\cite{borzacchiello2021fuzzing,even2023grayc}.

Besides functionality, another objective of debloating approaches is to enhance program security by removing unnecessary code, thereby reducing the attack surface~\cite{heo2018effective,xin2022studying}. However, while existing studies can reduce the attack surface to some extent~\cite{xin2022studying}, they often introduce more security issues than they mitigate (see \S\ref{sec:security}).

To maintain the functionality and security of debloated programs, we propose leveraging the strengths of LLMs by constructing a debloating framework that consists of documentation-guided test augmentation and multi-advisor-aided program debloating. For functionality, the test augmentation module utilizes the LLMs' semantic understanding to comprehend essential features with the aid of program documentation and generates sufficient tests to ensure coverage of desired features.
For security, the multi-advisor-aided debloating process ensures the security of the debloated program. It does this by evaluating the debloating suggestions and delineating potential consequences from advisors.

\subsection{Large Language Models}
Recent advancements in natural language processing have significantly improved the performance and adoption of LLMs. These advancements have enabled the scaling of LLMs to billions of parameters and the use of extensive training datasets containing billions of samples. LLMs are generally designed to be versatile, capturing knowledge from various domains. Consequently, they have garnered significant interest from researchers and have been widely applied to various downstream tasks, demonstrating remarkable capabilities across multiple aspects. 

For example, LLMs exhibit impressive abilities in understanding complex documentation~\cite{yuan2023evaluating,nam2024using}. Nam \etal~\cite{nam2024using} found that LLMs can assist developers by comprehending complex program documentation, interpreting code snippets, and even explaining key domain-specific terms. 
Building on their understanding capabilities, LLMs also demonstrate remarkable generative abilities across various domains, producing outputs ranging from natural language text~\cite{acharya2023llm,dong2023towards} to code~\cite{deng2023large,ryan2024code}, and even synthesizing strictly formatted test cases~\cite{chen2024chatunitest,xia2024fuzz4all}. For instance, LLM-based fuzzing has achieved 13\% to 76\% higher code coverage compared to traditional fuzzers~\cite{xia2024fuzz4all}.
In addition to understanding and generation, LLMs exhibit strong decision-making abilities, which is considered as a part of intelligence~\cite{loftus2020artificial,bruine2020decision}. Systems such as ``LLM-as-a-Judge" have demonstrated performance comparable to human judgment across a variety of tasks, underscoring their capacity to evaluate information and produce well-reasoned decisions~\cite{zheng2023judging,chang2024survey,huang2024empirical}.


Despite the benefits of LLMs, some researchers point out their limitations in specific code-related tasks, such as identifying failure-inducing test cases from similar tests~\cite{li2023nuances}. Additionally, as the test size increases, the ability of LLMs to memorize and process information diminishes~\cite{zhai2023investigating,kemker2018measuring}, limiting their ability to handle long inputs directly. 
Moreover, LLMs cannot autonomously complete complex tasks and need proper guidance.
Therefore, LLMs are not a one-size-fits-all solution and require carefully designed modules and workflows to achieve their full potential.

\section{Approach}
In this section, we present a motivating example to highlight key challenges of existing input-based program debloating approaches, followed by a comprehensive overview of \toolname. We then detail the implementation phases, including documentation-guided test augmentation, multi-advisor-aided program debloating, and validation.

\subsection{Motivation}
\begin{figure}
    \centering
    \begin{subfigure}[b]{0.45\textwidth}
        \begin{lstlisting}[language=C, breaklines=true, basicstyle=\ttfamily\normalsize]
int yyparse(parser_control *pc) {
  ...
  case_27:
    pc->time_zone =
        time_zone_hhmm(pc, HourOffset, MinOffset);
    goto switch_break;
  ...
}
        \end{lstlisting}
        \vspace{-4mm}
        \caption{The code snippet from {\tt date} program}
        \label{subfig:yyparse}
    \end{subfigure}
    \begin{subfigure}[b]{0.45\textwidth}
        \begin{lstlisting}[language=bash, basicstyle=\ttfamily\normalsize]
date -d "2023-01-01 12:00:00 -0500"
        \end{lstlisting}
        \vspace{-3mm}
        \caption{The unseen test during debloating}
        \label{subfig:time_zone}
    \end{subfigure}

    \caption{The {\tt yyparse} and corresponding inputs in {\tt date} program}
    \label{fig:enter-label}
\end{figure}
\begin{figure}
    \centering
    \begin{lstlisting}[language=C, basicstyle=\ttfamily\normalsize]
_Bool c_isspace(int c) {
  int tmp;
  if (c == 32) {
    tmp = 1;
  } else {
    if (c == 9) {
      tmp = 1;  // removed
    } else { ... }
  }
  return ((_Bool) tmp);
}
    \end{lstlisting}
    \caption{The code debloated by existing input-based techniques}
    \label{fig:moivation_vul}
\end{figure}
\label{subsec:motivation}

\begin{figure*}[htbp]
    \centering
    \includegraphics[width=0.85\linewidth]{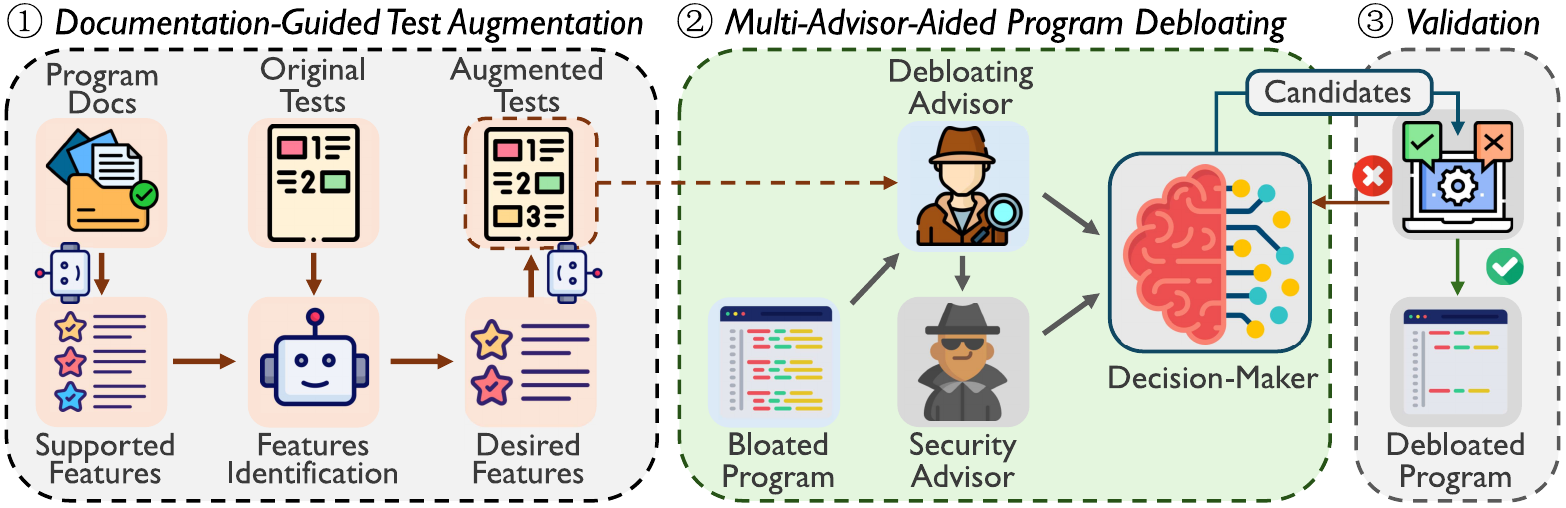}
    \caption{The workflow of \toolname}
    \label{fig:overview}
\end{figure*}

In the context of software debloating, the community-acknowledged has identified two primary objectives~\cite{heo2018effective,xin2022studying}: maintaining functionality and enhancing security. For functionality, the goal is to maintain program integrity by removing superfluous code while preserving the required features. For security, the objective is to enhance security by minimizing the attack surface through the elimination of vulnerable code segments, while also avoiding the introduction of new security issues during the debloating process. However, existing input-based program debloating techniques fall short of simultaneously achieving these two goals, as observed in prior approaches:

\noindent
\textbf{Challenge 1: Degradation of Program Functionality}. 
Eliminating code snippets associated with undesired features requires a precise specification of the program. However, creating accurate and comprehensive specifications for desired software features is challenging. Existing input-based debloating approaches rely on a constrained set of tests to identify the required code~\cite{xin2022studying,heo2018effective,qian2019razor}. These tests, however, can only cover a partial of code in terms of required features.
As a result, certain critical code segments may remain untriggered during the analysis, leading to their unintended removal and subsequent functionality loss.
For example, Figure~\ref{subfig:yyparse} shows a code snippet from the Unix {\tt date} program. The {\tt case\_27} block is designed to handle timezone offsets by normalizing them through {\tt time\_zone\_hhmm} and then assigning them to the relevant variable.
However, because no observed tests include a timezone offset, the entire {\tt case\_27} block is eliminated, causing the program to fail in cases requiring timezone specification, such as when adding a timezone offset (e.g., {\tt -0500}) at the end of a command, as illustrated in Figure~\ref{subfig:time_zone}.
To mitigate this limitation, existing approaches attempt either to preserve potentially useful code statements~\cite{qian2019razor,xin2020program} or to expand the tests through fuzzing~\cite{xin2022studying}. However, existing techniques cover only a small subset of the missing code, and fuzzing methods heavily rely on tools like AFL~\cite{AFL}, which struggle to generate valid tests that pass lexical analysis~\cite{bohme2016coverage,borzacchiello2021fuzzing}. In the case shown in Figure~\ref{subfig:yyparse}, the state-of-the-art technique, \textsc{CovF}, despite being equipped with AFL~\cite{AFL} to augment tests, leaves the entire {\tt case\_27} branch pruned, highlighting the limitations of fuzzing in addressing this issue.

\noindent
\textit{\bf Challenge 2: Compromised Program Security}. Current debloating approaches remove undesired code from the original program directly. However, this straightforward way can compromise program security.
For example, Figure~\ref{fig:moivation_vul} shows the function {\tt c\_isspace} from {\tt date}. This function checks if a given character {\tt c} (ASCII value) is a whitespace character and returns {\tt 1} if true (indicating space, tab, or another whitespace). In this example, the highlighted line ({\tt tmp = 1}) was removed by all investigated input-based techniques because it was not executed during testing (\ie no tab characters appeared in the tests). This debloating compromises the security of the program. Specifically, when the character {\tt c} is a tab, the value of {\tt tmp} becomes undefined, causing {\tt c\_isspace} to return a garbage value. This undefined behavior opens the door for attackers to exploit vulnerabilities, such as arbitrary code execution. Similar security risks may arise when the debloating process removes critical code (\eg null pointer checks or memory boundary checkers). However, none of the existing techniques take security into account during the debloating process.

To address these challenges, we propose an LLM-aided program debloating approach, \toolname, to address these challenges. We hypothesize that the effectiveness of debloating can be significantly enhanced by leveraging the capabilities of LLMs in understanding, generation, and decision-making. Our approach consists of two key modules:

1. \textbf{Documentation-Guided Test Augmentation (addressing Challenge 1)}: LLMs' ability to understand syntax and semantics allows them to generate tests that closely match users' actual behavior~\cite{jiang2024fuzzing,xia2023universal}. The capabilities of understanding and generation enable the creation of tests that accurately simulate real user behaviors. The process first utilizes LLMs' semantic understanding to comprehend necessary features through program documentation. It then generates comprehensive tests to cover these desired features for the subsequent debloating process.

2. \textbf{Multi-Advisor-Aided Program Debloating (addressing Challenge 2)}: Existing studies show that LLMs can serve as experts to incorporate diverse perspectives and make reasonable decisions~\cite{eigner2024determinants,chiang2024enhancing}, including detecting~\cite{yin2024multitask,lu2024grace} and repairing~\cite{pearce2023examining} vulnerabilities. Consequently, we introduce a multi-advisor-aided module where a debloater and a static analysis tool serve as advisors, and the LLM acts as the final decision-maker. The LLM is responsible for determining which parts of the code can be securely eliminated without introducing unexpected behaviors or vulnerabilities.

These two modules work together to preserve desired features while ensuring the debloated program maintains both its functional integrity and security.

\subsection{Workflow}

{The workflow of \toolname is outlined in Figure~\ref{fig:overview}, comprising three modules: documentation-guided test augmentation, multi-advisor-aided program debloating, and validation. 
In the documentation-guided test augmentation module, an LLM enhances tests by leveraging its understanding of syntax and semantics~\cite{jiang2024fuzzing,xia2023universal}.
Specifically, this module first extracts all supported features from its documentation. It then uses these extracted features along with given tests to identify the desired features. Finally, it generates tests targeting these desired features to increase their coverage.
In the multi-advisor-aided program debloating module, a debloater and a static analyzer serve as advisors, suggesting how to eliminate unnecessary code while ensuring functionality and security. These suggestions are evaluated by an LLM-based decision-maker.
During the validation stage, the debloated program candidates are compiled and tested against a suite of test cases. Any compilation errors or failed test cases are fed back into the program debloating component. This iterative process continues until the debloated programs either pass all test cases or reach the iteration limit.}

\subsection{Documentation-Guided Test Augmentation}
{In program debloating, users are required to provide a specification that describes the desired features~\cite{heo2018effective,xin2022studying}. In input-based debloating, the given tests act as the specification for the desired features. This implies that for each desired feature, there is at least one corresponding test case. Any code snippets that remain unexecuted are likely to be eliminated, as they are not invoked during the execution of these tests.
Therefore, the quality of the debloated program is directly influenced by both the quality and quantity of the tests. To enhance both, we construct a documentation-guided test augmentation component to improve the quality and diversity of the provided tests.

This component consists of three stages: documentation understanding, desired features identification, and test generation. The {\bf documentation understanding} focuses on analyzing the program's functionality as described in its documentation to identify all supported features, reflecting the full functionality the software is capable of. However, users typically require only a specific subset of these features rather than the full range of capabilities.
To address this distinction, the {\bf desired features identification} is designed to narrow down the supported features to those that align with the user's actual requirements, referred to as the {\em desired features}. This is achieved by analyzing the tests provided by the user, which serve as a concrete representation of the required features. By cross-referencing the features extracted from the documentation with those exercised in the tests, this stage determines which features are truly required. 
Finally, the {\bf test generation} stage uses the desired features and generates additional tests when provided with existing tests as examples to thoroughly cover the desired features. {\update Note that if the documentation of the desired features is missing, the test generation stage would use the user-provided tests to generate the tests.}


\begin{figure}
 \prompt{Prompt 1: Documentation Understanding}{1}{
You are tasked with analyzing the functionalities of a program based on its documentation. Your goal is to extract the key features, command-line options, and corresponding use cases.\\

\textbf{\# Task}\\
For the provided program documentation, perform the following steps:

\hspace{10pt}-Identify Program Features: ...

\hspace{10pt}-Extract Command-Line Options:...

\hspace{10pt}-Provide Usage Examples:...
\\

\textbf{\# Input}\\
The following is the program documentation:\\
\{{\footnotesize DOCUMENTATION}\}
\\

\textbf{\# Output}\\
For each feature identified:

\hspace{10pt}-Concise description of the feature.

\hspace{10pt}-All relevant command-line options. 

\hspace{10pt}-Concrete examples of how the feature is used.

}
\vspace{-2mm}
\end{figure}

\subsubsection{Documentation Understanding}
This phase aims to understand the functionality of the program and summarize the features it supports. To this end, we provide the documentation of the program, which contains descriptions of its functionalities and supported features. Specifically, for Unix programs, we obtain the documentation using the {\tt man} command to collect manual pages. For other programs, we invoke the help function by passing {\tt --help} or {\tt -h} to obtain documentation. 
The documentation understanding prompt template is shown in Prompt~\ref{prompt:1}.
Specifically, we instruct the LLM to extract the supported features from the documentation. We then specify the required information for each feature, including its description, command-line option, and use case. The extracted features, along with the tests, are then sent to the next stage to identify the desired features.

\subsubsection{Desired Features Identification}
After extracting all supported features from the documentation, we focus on retaining only the required ones. To achieve this, we instruct the LLM to identify the truly necessary features based on the extracted information and observable tests.
The prompt for this stage is as shown in Prompt~\ref{prompt:2}.
Initially, the LLM is instructed to identify features relevant to the given tests. We then provide the extracted feature descriptions from the documentation understanding stage, along with the given tests. Once the desired features are identified, their descriptions, command-line options, and examples are forwarded to the test generation stage. Notably, the provided tests are assigned as examples for their corresponding features.
\begin{figure}
    \centering
    \prompt{Prompt 2: Desired Features Identification}{2}{
You are given a list of feature descriptions for the program \{{\footnotesize PROGRAM NAME}\}, along with a list of tests. Your task is to identify which features each test relates to, based on the descriptions provided.\\

\textbf{\# Feature Descriptions}\\
\{{\footnotesize FEATURES}\}\\

\textbf{\# List of Tests}\\
The tests provided are commands or arguments that invoke the desired features of the program as follows:
\{{\footnotesize TESTS}\}
\\

\textbf{\# Output}\\
List the identified related features, including their descriptions, command line options and example tests. Include the provided tests if they relate to the identified features.

}
\vspace{-2mm}
\end{figure}

\subsubsection{Test Generation}
In this stage, we utilize the identified features and the list of tests to generate additional tests, aiming to cover the required functionality as thoroughly as possible. 
The prompt for this stage is shown in Prompt~\ref{prompt:3}.
Specifically, the LLM is first instructed to generate tests based on the feature descriptions. It then outputs the tests directly associated with the given features. These generated tests constitute the final output of the tests augmentation component and serve as the foundation for the subsequent program debloating component.
\begin{figure}
    \prompt{Prompt 3: Test Generation}{3}{
You are given a set of features with their descriptions, command-line options, and use cases for program \{{\footnotesize PROGRAM NAME}\}. Your task is to generate additional tests for each feature that cover a wide range of possible scenarios. \\

\textbf{\# Feature Descriptions}\\
Feature 1:

\hspace{10pt}-Feature Description: ...

\hspace{10pt}-Command Line Options: ...

\hspace{10pt}-Use Cases: ...

...

\textbf{\# Output}\\
Output the generated tests directly without explanation.
}
\vspace{-1mm}
\end{figure}



\subsection{Multi-Advisor-Aided Program Debloating}

After the test augmentation phase, we construct a multi-advisor-aided program debloating module, which employs a neuro-symbolic pipeline to enhance software security during the debloating process. The workflow of this module is detailed in Algorithm~\ref{alg:program_debloating}.
Overall, two advisors are employed:  
\begin{itemize}[leftmargin=*]
    \item A \textbf{debloating advisor} processes the tests augmented by the documentation-guided test augmentation module to identify potentially unused and unnecessary code. This advisor provides debloating suggestions, denoted as $D_{sug}$.
    \item A \textbf{security advisor}, which identifies potential security risks (\ie $S_{sug}$) that may arise if the debloating suggestions $D_{sug}$ are applied.
\end{itemize}

The core process iterates over each function in the original program (line 6), aiming to replace it with a debloated version. For each function, the decision-maker evaluates the debloating and security suggestions to generate a candidate function (line 9). This candidate is validated by applying it to the program and testing its correctness (lines 10-11). If the candidate passes validation, it is added to a pool of accepted debloated functions (line 13); otherwise, the failed attempt informs subsequent iterations (line 15). Finally, the accumulated debloated functions are applied to the original program, producing the debloated version (line 16).

\begin{algorithm}[]
\small
\caption{{\small Multi-Advisor-Aided Program Debloating}}
\label{alg:program_debloating}
\KwIn{$T_d$: Tests for validation, $T_{aug}$: Augmented tests, $P_{ori}$: original bloated program}
\KwOut{$P_{deb}$: Debloated program}

\textbf{Step 1: Generate Suggestions:}\\
$D_{sug} \gets$ DebloatingAdvisor($T_{aug}$)\\
$S_{sug} \gets$ SecurityAdvisor($D_{sug}$) 

\textbf{Main Algorithm:}\\
$cand \gets \emptyset$ \tcp{Store debloated functions}
\ForEach{$f_{ori}$ in $P_{ori}$}{
    $fail \gets \textbf{None}$ \tcp{Track failed attempts}
    \For{$i \gets 1$ \textbf{to} $max\_iter$}{
        $f_{deb} \gets$ DecisionMaker($D_{sug}, S_{sug}, f_{ori}, fail$)\\
        $P_{tmp} \gets$ ApplyDebloating($f_{deb}, P_{ori}$)\\ 
        $corr \gets$ Valid($T_d, P_{tmp}$)\\
        \If{$corr = \textbf{True}$}{
            $cand \gets cand \cup \{f_{deb}\}$\\
            \textbf{break}
        }
        $fail \gets f_{deb}$\tcp*{Update failed attempt}
    }
}

$P_{deb} \gets$ ApplyDebloating($cand, P_{ori}$)\\
\Return $P_{deb}$
\end{algorithm}


\subsubsection{Debloating Advisor}
The debloating advisor provides suggestions on which statements may be eliminated based on coverage and static analysis. Specifically, given a program $P$ and a set of tests $T$, the advisor first instruments $P$ and executes it against $T$ to identify all statements exercised by the tests. Unexecuted statements are marked as candidates for deletion. We implement this advisor using tools such as llvm-cov~\cite{llvm_cov} and gcov~\cite{gcc_gcov} for instrumentation, and Clang~\cite{clang_llvm} to construct an abstract syntax tree (AST). The advisor collects the locations of all program statements by traversing the AST and analyzes the coverage report generated by the instrumentation tools, comparing it with the statement locations to identify unexecuted statements.

Furthermore, the advisor also identifies code that, although executed, is unnecessary. We primarily focus on four categories of unnecessary code, following prior work~\cite{xin2022studying}: (1) unused variable declarations, (2) null statements, (3) side-effect-free conditions, and (4) unused label statements. To detect unused variable declarations, we perform a def-use chain analysis and mark the declaration statements of unused variables as deletable. For null statements and side-effect-free conditions, we traverse the AST and apply predefined rules (e.g., a condition statement with an empty body is marked as a null statement). For unused label statements, we record defined but unused labels during the AST traversal.

\subsubsection{Security Advisor}
\label{subsec:securityAdvisor}
Previous studies~\cite{charoenwet2024empirical,mehrpour2023can} have shown that static analysis tools are valuable for helping developers identify security vulnerabilities and enforce coding standards. In this study, we utilize static analysis to detect potential insecure behaviors introduced during the debloating process, such as null pointer dereferences and buffer overflows. Specifically, we employ CodeChecker~\cite{CodeChecker2024}, which integrates several analyzers (\eg Cppcheck~\cite{Cppcheck2024} and Infer~\cite{Infer2024}) and has demonstrated strong performance in detecting vulnerabilities, as highlighted in prior empirical research~\cite{charoenwet2024empirical}.
Despite their utility, static analysis tools are often hindered by a high false-positive rate, where secure code is incorrectly flagged as vulnerable~\cite{kang2022detecting,kharkar2022learning}.
To address this challenge, we adopt two strategies: differential analysis and the integration of an LLM as a decision-maker. First, we apply differential analysis by performing static analysis on both the original and debloated versions of the program. By comparing the detected vulnerabilities in each version, we can identify the vulnerabilities that have been eliminated as well as those that have been introduced during the debloating process.

Formally, for a program with the original version \( P_{\text{ori}} \) and the debloated version \( P_{\text{deb}} \), we first analyze the differences between the two programs by analyzing the deleted lines. Let \( D \) denote the set of differences, representing the lines of deleted code. Specifically, \( D \) can be expressed as:
\begin{equation*}
D = \{ (l_{\text{ori}}, \_) \mid l_{\text{ori}} \in P_{\text{ori}}, l_{\text{ori}} \notin F_{\text{deb}} \}
\end{equation*}
Next, we perform static analysis on both programs, \( P_{\text{ori}} \) and \( P_{\text{deb}} \), to obtain the sets of vulnerabilities. Let the static analysis function be denoted as \( SA \); then we have:
\begin{equation*}
V_{\text{ori}} = SA(P_{\text{ori}})
\end{equation*}
\begin{equation*}
V_{\text{deb}} = SA(P_{\text{deb}})
\end{equation*}
where \( V_{\text{ori}} \) and \( V_{\text{deb}} \) are the sets of vulnerabilities before and after debloating, respectively. Consequently, the vulnerabilities eliminated during debloating, \( V_{\text{elim}} \), are given by:
\begin{equation*}
V_{\text{elim}} = \{ v \in V_{\text{deb}} \mid \text{line}(v) \in D \}
\end{equation*}
where \( \text{line}(v) \) denotes the line of code where vulnerability \( v \) is located. 

Similarly, the vulnerabilities introduced, \( V_{\text{new}} \), can be defined as:
\begin{equation*}
V_{\text{new}} = \{ v \in V_{\text{deb}} \mid v \notin (V_{\text{ori}} \setminus V_{\text{elim}}) \}
\end{equation*}
The eliminated vulnerabilities serve as metrics for further evaluation, while the introduced vulnerabilities are fed into the subsequent LLM-based decision-maker.
Second, inspired by recent studies~\cite{li2024enhancing,li2024llm} that demonstrate how LLMs can reduce false-positive rates in static analysis, we incorporate an LLM to evaluate the validity of flagged vulnerabilities. The LLM also assists in deciding which code segments can be safely debloated and the appropriate debloating method. Further details on the decision-making process are discussed in \S\ref{subsubsec:decition_maker}.

\subsubsection{Decision-Maker}
\label{subsubsec:decition_maker}

\begin{figure}
    \centering
    \begin{lstlisting}[language=C,
        basicstyle=\ttfamily\normalsize]
_Bool c_isspace(int c) {
  int tmp;
  if (c == 32) {
    tmp = 1;
  } else {
    if (c == 9) {
//      tmp = 1;
    } else { ... }
  }
  return ((_Bool) tmp); /*Uninitialized Value: The value read from tmp was never initialized.*/
}
    \end{lstlisting}
    \caption{The code along with integrated suggestions}
    \label{fig:suggestions_case}
\end{figure}
We employ an LLM as a decision-maker to handle suggestions for debloating and security enhancement. The decision-making process assists in identifying which parts of the code should be debloated and determining the appropriate debloating method. The most straightforward approach involves sequentially presenting all relevant information and allowing the LLM to infer relationships between the various data points.

However, a key challenge arises when the LLM processes the input from security advisors, such as messages like ``pointer {\tt P} at line 2366 could be null." LLMs are known to struggle with precise counting tasks, as highlighted by researchers from Google~\cite{yehudai2024can}. This limitation makes it difficult for the LLM to handle exact line references or numerical data in code. Additionally, we must minimize redundancy in the provided information, as the performance of LLMs tends to decline with increasing input length~\cite{levyetal2024task}. 

To mitigate these challenges, we use the debloated program as the fundamental context and integrate all external information, such as security advisor suggestions, directly into the original code. Figure~\ref{fig:suggestions_case} demonstrates how this external information is incorporated. Specifically, the debloating advisor annotates statements (\eg {\tt tmp=1}) that are recommended for debloating based on the analysis of coverage and static information, while security suggestions are added as comments at the end of problematic statements (\eg at the end of the return statement to indicate that this elimination may trigger an uninitialized value for {\tt tmp}). Note that for some messages containing an error line number, we append these messages at the end of the indicated line and replace the phrase ``at line [NUM]'' with ``at this line''.

Once suggestions for debloating the target code have been extracted, we construct a prompt for the LLM using the structure as shown in Prompt~\ref{prompt:4}.
\begin{figure*}
\prompt{Prompt 4: Decision-Maker}{4}{
You are a decision-maker tasked with identifying code that can be securely removed because it is unrelated to the required features. A debloating advisor will suggest code snippets for elimination, while a security advisor will provide advice about vulnerabilities that may be introduced when applying the debloating recommendations.\\

\textbf{\# Goal}\\
As a senior software engineer specializing in code debloating, your objective is to eliminate statements irrelevant to the required features while ensuring no vulnerabilities are introduced. You must then refactor the function to ensure both functionality and security.\\

\textbf{\# Input Format Definition}\\
A \{{\footnotesize LANGUAGE}\} language function, potentially bloated, will be provided with annotations: 

\hspace{10pt}- The debloating advisor will indicate which statements is suggested for removal by annotating them, as it is deemed unrelated to the required features.

\hspace{10pt}- The block comment at the end of each statement outlines concerns raised by the security advisor, highlighting potential vulnerabilities if the suggested debloating is applied.\\

\textbf{\# Input}\\
\{{\footnotesize CODE WITH SUGGESTIONS}\}\\

\textbf{\# Failure Case} (Optional)\\
Here is a debloated function from the previous decision-maker:
\{{\footnotesize DEBLOTED FUNCTION}\}.

This function raises the following compiler errors:
\{{\footnotesize COMPILE ERROR}\}.

This function fails the following test cases:
\{{\footnotesize FAILED TEST CASES}\}.\\
The expected output is: \{{\footnotesize EXPECTED OUTPUT}\}, but the actual output is: \{{\footnotesize ACTUAL OUTPUT}\}.
\\

\textbf{\# Output}\\
First, analyze the code to identify statements that can be securely removed without introducing vulnerabilities, guided by recommendations from debloating and security advisors. For failure cases, refer to them to guide debloating while ensuring no exposed issues remain.\\

}
\end{figure*}
The prompt begins with task-specific instructions, followed by the goal description ({\#Goal}).
Next, the format definitions for debloat and security suggestions are provided. We then present the code to be debloated, complete with integrated suggestions (\#Input). Finally, we outline the expected output of the task (\#Output), instructing the LLM to first analyze the inquired statements and then produce the debloated code in the given format.

\subsection{Validation}
Each generated function undergoes a separate validation process using the test cases, integrated into the original program one at a time.
Note that the tests in this phase are used only for debloating. The tests used for evaluation are not utilized here, ensuring fair comparison, as illustrated in \S\ref{subsec:testcase}.
Specifically, we replace the corresponding function in the program with its newly debloated version, then compile and execute the associated test cases. If a generated function fails to compile, or compiles but fails some test cases, the compiler's error message or information about the failed test cases (detailed in \S\ref{subsubsec:decition_maker}) is fed back to the decision-maker. This process is repeated until the function is successfully debloated or the maximum number of iterations (discussed in \S\ref{subsec:hyperpara}) is reached.

\section{Experiment Settings}
\label{sec:exp_setting}
\subsection{Baselines}
We select the following state-of-the-art techniques as baselines: 

\begin{itemize}[leftmargin=*]
\item \textsc{Chisel} is a source code debloating tool that uses the Markov decision process for delta debugging to eliminate unnecessary parts of a program.

\item \textsc{Debop} formulates the process of debloating a program as a multi-objective optimization problem. It performs stochastic optimization to minimize program size and attack surface while maximizing generality.

\item \textsc{Razor} is a binary-level approach that first constructs a CFG through program execution traces using given tests. It then utilizes pre-defined heuristics to expand the CFG to include some related but not executed code, generating the debloated program.

\item \textsc{Cov}, \textsc{CovF} and \textsc{CovA}. \textsc{Cov} first collects coverage against tests and then removes all unexecuted statements. Compared to \textsc{Cov}, \textsc{CovF} uses fuzzing to produce more tests and includes the covered statements from additional tests that crash the program. Unlike \textsc{CovF}, \textsc{CovA} directly identifies functions that may be feature-related and preserve them entirely in the final debloated program by calculating the score of each function through heuristics.

\item Cov-Perf. To evaluate the extent of code redundancy and its potential for removal, we grant the debloater access to all tests, including those typically reserved for evaluation. This comprehensive approach allows for optimal debloating by providing access to all input scenarios. By establishing this baseline, we can effectively assess how much code is redundant for preserving desired features and whether other approaches are excessively reducing code.
We implement this methodology based on \textsc{Cov} and designate it as Cov-Perf.
\end{itemize}

\subsection{Evaluation Metrics}
\label{susec:metrics}
In this study, we assess the effectiveness of various approaches based on reduction, correctness, robustness, and harmonized score by measuring the harmonic mean of correctness and robustness. Additionally, we evaluate the security hardening achieved by measuring the vulnerabilities reduced and introduced by debloaters, as security hardening is a key objective of program debloating~\cite{heo2018effective}.

\subsubsection{Reduction}
For a program $P$ and its debloated version $P'$, we assess the reduction of program $P'$ in terms of three aspects: program size, memory usage, and attack surface. Specifically, given $P$ and $P'$, the reduction in program size is computed as follows:
\begin{equation*}
    size\_red = \frac{size(P)-size(P')}{size(P)}
\end{equation*}
\noindent
where $size(\cdot)$ measures the program size by calculating the number of statements in the given program. 

Similarly, memory usage reduction is calculated as:
\begin{equation*}
    mem\_red = \frac{mem(P)-mem(P')}{mem(P)}
\end{equation*}
\noindent
where $mem(\cdot)$ measures the memory usage by calculating the number of bytes in the executable memory region using the tool provided by \textsc{Razor}.

The reduction of the attack surface, given variables $P$ and $P'$, is computed as follows:
\begin{equation*}
    atk\_red = \frac{atk(P)-atk(P')}{atk(P)}
\end{equation*}
\noindent
where $atk(\cdot)$ measures the attack surface of a given program by calculating the number of return-oriented programming (ROP) gadgets it contains. An ROP gadget is a sequence of machine instructions that typically ends with a return instruction. An attacker could potentially use it to overwrite a gadget’s return address, divert the control flow, and execute malicious code~\cite{shacham2007geometry}.
For the calculation of $size(\cdot)$, $mem(\cdot)$, we reuse the tool provided by \textsc{Chisel}. For the calculation of $atk(\cdot)$, we reuse the ROPgadget tool~\cite{ROPgadget2022} to count ROP gadgets following previous studies~\cite{heo2018effective,qian2019razor}.



\subsubsection{Correctness}
Given a program $P$ and its debloated version $P'$ generated based on a set of tests $T$, we focus on the functional correctness of the debloated program. These measures assess the extent to which $P$ can behave correctly for additional tests not included in $T$. To measure the correctness of $P$, we leverage a distinct set of tests $T'$ ($T' \cap T = \emptyset$) and evaluate the proportion of cases from $T'$ where $P'$ operates correctly. Formally, for correctness is computed as follows:
\begin{equation*}
    cor(P',T')=\frac{\sum_{t\in T'} P'(t)=P(t)}{\left| T'\right|}
\end{equation*}
\noindent
where $P(\cdot)$ and $P'(\cdot)$ denote the outputs of $P$ and $P'$, and $\left| T'\right|$ represents the number of test cases in $T'$.

\subsubsection{Robustness}
Similar to correctness, robustness measures the extent to which $P'$ can execute reliably for tests not included in $T$. Formally, robustness is computed as follows:
\begin{equation*}
    rob(P',T')=\frac{\sum_{t\in T'} robust(P'(t))}{\left| T'\right|}
\end{equation*}
\noindent
where $robust(\cdot)$ indicates whether the program terminates without crashing (\eg no segmentation fault). Unlike the correctness measure, where $T'$ is gathered from real-world scenarios, $T'$ in robustness is generated by mutating $T'$ using a fuzzer, following prior work~\cite{xin2022studying}. The rationale for using fuzzing to generate tests is that more tests can evaluate robustness more comprehensively by leveraging a greater number of test cases. The details are illustrated in \S\ref{subsec:testcase}.

\subsubsection{Harmonized Score}
Correctness and robustness are equally critical for a debloated program to be practical and reliable. To provide a balanced measure for evaluating the performance of debloating approaches, we define harmonized score by calculating the harmonic mean of correctness and robustness as follows:
\begin{equation*}
    harmonized\_score(cor,rob)=\frac{2\cdot cor\cdot rob}{cor+rob}
\end{equation*}
\noindent
where $cor$ represents the correctness score, and $rob$ denotes the robustness score. A high score indicates a balanced and simultaneous achievement of both correctness and robustness. Thus, the harmonized score provides a comprehensive assessment of an approach's overall performance in terms of correctness and robustness.


\subsubsection{Security Hardening}
To assess the efficacy of tools in terms of security hardening, we utilize CodeChecker~\cite{CodeChecker2024} to detect vulnerabilities before and after debloating. Specifically, we focus on the number of vulnerabilities eliminated through debloating versus those introduced by the process. The formal definitions of eliminated and introduced vulnerabilities are provided in \S\ref{subsec:securityAdvisor}.


Additionally, CodeChecker categories detected vulnerabilities into Low, Medium, and High severity levels based on their impact, ease of exploitation, and other factors. We use the severity ratings reported by CodeChecker to analyze the number of vulnerabilities across different severity levels that are eliminated and introduced during the debloating process.

\subsection{Benchmark}
\label{sec:benchmark}
\subsubsection{Programs}
\begin{table}[!t]
  \centering
  \caption{The statistic of benchmarks.}
  \resizebox{0.97\linewidth}{!}{
    \begin{tabular}{c|lrrrr}
    \toprule
    Benchmark & Program & \#Line & \#Func & \#Stmt & \#Tests \\
    \hline
    \multirow{11}[3]{*}{Util} & bzip2-1.0.5 & 11,782 & 97    & 6,154 & 59 \\
          & chown-8.2 & 7,081 & 122   & 3,765 & 111 \\
          & date-8.21 & 9,695 & 78    & 4,228 & 174 \\
          & grep-2.19 & 22,706 & 315   & 10,977 & 145 \\
          & gzip-1.2.4 & 8,694 & 91    & 4,049 & 81 \\
          & mkdir-5.2.1 & 5,056 & 43    & 1,804 & 50 \\
          & rm-8.4 & 7,200 & 135   & 3,835 & 84 \\
          & sort-8.16 & 14,264 & 233   & 7,805 & 117 \\
          & tar-1.14 & 30,477 & 473   & 13,995 & 84 \\
          & uniq-8.16 & 7,020 & 65    & 2,086 & 72 \\
\cmidrule{2-6}          & Total & 123,975 & 1,652  & 58,698 & 977 \\
    \hline\hline
    \multirow{16}[4]{*}{SIR} & bash-2.05 & 58,319 & 1,003 & 27,646 & 1,061 \\
          & flex-2.5.4 & 15,518 & 162   & 6,704 & 670 \\
          & grep-2.4.2 & 16,203 & 131   & 8,437 & 806 \\
          & gzip-1.3 & 8,882 & 97    & 4,287 & 213 \\
          & make-3.79 & 26,118 & 248   & 12,901 & 1,832 \\
          & sed-4.1.5 & 18,866 & 247   & 9,179 & 370 \\
          & space & 8,215 & 136   & 4,376 & 13,549 \\
          & vim-5.8 & 136,531 & 1,699 & 66,080 & 975 \\
          & printtokens & 1,069 & 18    & 396   & 4,073 \\
          & printtokens2 & 824   & 19    & 341   & 4,058 \\
          & replace & 938   & 21    & 416   & 5,542 \\
          & schedule & 537   & 18    & 211   & 2,650 \\
          & schedule2 & 604   & 16    & 238   & 2,710 \\
          & tcas  & 382   & 9     & 162   & 1,608 \\
          & totinfo & 586   & 7     & 265   & 1,052 \\
\cmidrule{2-6}          & Total & 293,592 & 3,831  & 2,029 & 41,169 \\
    \bottomrule
    \end{tabular}%
  }
  \label{tab:dataset}%
\end{table}%

In our evaluations, we used two sets of programs following the previous study~\cite{xin2022studying}: the {\tt Util} benchmark, which contains ten utility programs, and the {\tt SIR} benchmark~\cite{sir2022bench}, which contains 15 programs. The {\tt Util} benchmark has been widely used in earlier debloating studies~\cite{xin2022studying,qian2019razor,heo2018effective}. The {\tt SIR} benchmark is sourced from the Software-artifact Infrastructure Repository~\cite{sir2022bench}, designed to support controlled experimentation with program analysis and software testing techniques.

Table~\ref{tab:dataset} provides statistics on two benchmarks, detailing the size of programs by lines of code (\#Line), number of functions (\#Func), number of statements (\#Stmt), and number of tests (\#Tests). The program sizes vary significantly, with {\tt tcas} at 382 lines and {\tt vim} at 136,531 lines.
In the {\tt Util} benchmark, {\tt tar-1.14} is the largest program, comprising over 400 functions and 10,000 statements. The number of tests is a critical factor for evaluating a debloated program's correctness across various scenarios. On average, the {\tt Util} benchmark suite includes 97.7 tests per program, facilitating a comprehensive assessment of correctness and reliability after debloating.
Referring to the {\tt SIR} benchmark, the size of programs and the number of tests vary significantly across the included programs. The largest program (\ie {\tt vim-5.8}) contains over 136,000 lines of code, whereas {\tt tcas} consists of only nine functions. This diversity in program sizes and complexities provides a valuable basis for evaluating the performance of debloating approaches in handling a wide range of program characteristics.


\subsubsection{Tests}
\label{subsec:testcase}
We categorize the tests of a program into two subsets on a scale of 1:9, reflecting the debloating scenario where users can only provide a limited number of tests as specifications for the desired features~\cite{xin2022studying}: $T_{d}$ for debloating and $T_{e}$ for evaluation. The tests in $T_{d}$ are observable to the debloater, whereas the tests in $T_{e}$ serve as test cases to evaluate whether the debloated program produces the same output as the original program. 
Since the primary objective of debloating is to retain code snippets corresponding to the desired features inferred from the given tests, we split $T_{d}$ and $T_{e}$ based on the similarity between the tests in the two sets, following the methodology of previous studies~\cite{xin2022studying}. In other words, the tests in $T_{d}$ and $T_{e}$ are related but do not overlap, such as tests that invoke the same features. To ensure a fair comparison, we utilize the same test-splitting methodology as employed in the previous study~\cite{xin2022studying}. 



In addition to correctness, we also evaluate the robustness of the investigated approaches. Unlike the evaluation of correctness, which assumes that a debloated program is correct if its output matches the original program's output, the evaluation of robustness checks whether a program crashes or fails to terminate. Specifically, we measure robustness by running AFL~\cite{Afl2022}. The fuzzer tests the debloated program with randomly generated command-line tests and input files (when necessary) based on the testing tests. We set a five-hour time threshold for fuzzing each debloated program, following the setting of a previous study~\cite{xin2022studying}.
To determine whether a program has crashed, we check for specific exit codes (131–136 and 139) that indicate abnormal termination, such as segmentation faults. Non-termination is determined by a ten-second time limit per test case, with programs exceeding this limit considered non-terminated.

\subsection{Research Questions}
In our study, we seek to answer the following research questions (RQs):

\noindent
{\bf RQ1: Compare with state-of-the-art techniques.} How does \toolname perform compare with existing software debloating techniques?

\noindent
{\bf RQ2: Security of debloated programs.} How secure is the debloated program compared to existing program debloating techniques?  


\noindent
{\bf RQ3: Ablation study.} What are the contributions of the major components of \toolname?

\subsection{Implementation Details}
Within the workflow of \toolname, we use DeepSeek-V2.5~\cite{DeepSeek2024} as the LLM backend in this paper. It demonstrates outstanding performance in code generation and instruction compliance, comparable to GPT-4, but offers a more preferential price of approximately 1 million input tokens for \$0.14, while GPT-4 costs \$2.50 per million tokens. To effectively utilize the inherent randomness of LLMs, we set the temperature to 0.5. We limit the maximum number of iterations to three per function. The impact of temperature and maximum iteration number is discussed in \S\ref{subsec:hyperpara}. 
All other configurations are kept at their default values.

\section{Experiment Results}
\label{sec:exp_results}
\begin{table}[htbp]
  \centering
  \caption{The performance of different approaches on benchmarks.}
  \resizebox{1\linewidth}{!}{
    \begin{tabular}{c|c|c|c|c|ccc}
    \toprule
    \multirow{2}[2]{*}{Benchmark} & \multirow{2}[2]{*}{Tool} & \multirow{2}[2]{*}{Cor$^{\dagger}$} & \multirow{2}[2]{*}{Rob$^{\dagger}$} & \multirow{2}[2]{*}{HS$^{\dagger}$} & \multicolumn{3}{c}{Reduction} \\
\cmidrule{6-8}          &   &   &  & & Size  & Mem$^{\ddagger}$ & Atk$^{\ddagger}$ \\
    \midrule
    \multirow{7}[1]{*}{Util} & \textsc{Chisel} & 0.55 & 0.61 & 0.58 & 0.71 & 0.65 & 0.36 \\
          & \textsc{Debop} & 0.54 & 0.60 & 0.56 & {\bf 0.76} & 0.67 & 0.39 \\
          & \textsc{Razor} & 0.69 & 0.75 & 0.72 & - & {\bf 0.70} & {\bf 0.51} \\
          & Cov   & 0.57 & 0.57 & 0.57 & 0.74 & 0.65 & 0.33 \\
          & CovA  & 0.73 & 0.73 & 0.73 & 0.65 & 0.57 & 0.26 \\
          & CovF  & 0.61 & {\bf 0.98} & 0.75 & 0.71 & 0.62 & 0.31 \\
          & \toolname & {\bf 0.96} & 0.93 & {\bf 0.94} & 0.54 & 0.52 & 0.20 \\
          \cmidrule{2-8}
          & Cov-Perf & 0.98 & 1.00 & 0.99 & 0.38 & 0.45 & 0.18 \\
    \hline\hline
    \multirow{7}[1]{*}{SIR} & \textsc{Chisel} & 0.60 & 0.58 & 0.59 & 0.33  &  0.26 & 0.07 \\
          & \textsc{Debop} & 0.58 & 0.54 & 0.56 & {\bf 0.47} & 0.37 & 0.16 \\
          & \textsc{Razor} & 0.88 & 0.86 & 0.87 & -	& {\bf 0.55} & {\bf 0.24} \\
          & Cov   & 0.64 & 0.58 & 0.60 & 0.40 & 0.24 & 0.05 \\
          & CovA  & 0.68 & 0.61 & 0.64 & 0.39 & 0.23 & 0.05 \\
          & CovF  & 0.71 & {\bf 0.99} & 0.83 & 0.39 & 0.23 & 0.08 \\
          & \toolname & {\bf 0.95} & 0.94 & {\bf 0.94} & 0.31 & 0.26 & 0.04 \\
          \cmidrule{2-8}
          & Cov-Perf & 0.98 & 0.99 & 0.98 & 0.22 & 0.25 & 0.02 \\
    \bottomrule
    \end{tabular}%
    }
    {\footnotesize
    \begin{flushleft}
    $^{\dagger}$ Cor, Rob, and HS stand for correctness, robustness, and harmonized score, respectively.\\
    $^{\ddagger}$ Mem and Atk represent the memory and attack surface, respectively.
    \end{flushleft}
    }
  \label{tab:cor_based_perf}%
\end{table}%

\subsection{RQ1: Compare with state-of-the-art techniques}
In this research question, we assess the effectiveness of \toolname and baselines in terms of correctness, robustness, harmonized score, and reduction, as outlined in Section~\ref{susec:metrics}. Table~\ref{tab:cor_based_perf} presents results for all approaches. Note that size reductions for \textsc{Razor} are not available, as its reduction is measured by comparing the number of statements at the source level before and after debloating, whereas \textsc{Razor} performs debloating at the binary level.

Overall speaking, \toolname demonstrates notable advantages, particularly in the correctness and robustness dimensions. It achieves a correctness score (Cor) of 0.96 and a robustness score (Rob) of 0.93 on {\tt Util} benchmark, the highest among the listed tools for correctness and nearly the highest for robustness. Similar high scores were also achieved for the {\tt SIR} benchmark.
This strong performance in correctness indicates that \toolname can effectively retain the correct functionality of the program while debloating, even under varying conditions and usage scenarios. This performance closely approximates the ground-truth (\ie {\tt Cov-Perf}), which uses all available tests (including the unseen tests) for debloating. Note that the correctness of {\tt Cov-Perf} is not exactly 1 due to some test cases producing time-dependent outputs, which cause inconsistencies between the original and debloated programs. 
Compared with state-of-the-art technique, \textsc{CovA}, achieves 0.73 and 0.68 on correctness scores on {\tt Util} and {\tt SIR}, respectively. This indicates that, in about 30\% of test cases, the debloated program output produced by \textsc{CovA} is inconsistent with the original program.
The fuzzing-based approach \textsc{CovF}, despite generating numerous external tests through fuzzing, achieves correctness scores of only 0.61 and 0.71 on two benchmarks. We hypothesize that while fuzzing generates a high volume of tests, these tests often include irregular cases that exercise non-core logic~\cite{even2023grayc,borzacchiello2021fuzzing}, making it challenging to adequately cover essential program features. 

The harmonized score dimension provides a comprehensive measure of how well necessary functions are preserved with robustness. A high score in this dimension indicates a balanced and simultaneous achievement of both correctness and robustness. Specifically, \toolname achieves a harmonized score of 0.94, representing improvements of 28.8\% and 25.3\% compared to CovA and CovF, respectively. This result demonstrates that \toolname performs well in both correctness and robustness. In contrast, CovF, which leverages fuzzing techniques, achieves an impressive robustness score of 0.98 but only attains a correctness score of 0.61, highlighting its imbalance in correctness and robustness.

Although \toolname shows inspiring results in correctness, robustness, and harmonized score, its focus on preserving correctness appears to limit its performance in the reduction metrics. While some other tools, such as \textsc{Debop}, achieve significant reductions in both size and memory (with scores of 0.76 and 0.67 in {\tt Util}, respectively), \toolname shows moderate reductions in these areas, with size and memory scores of 0.54 and 0.52, respectively. This moderate reduction suggests that \toolname may retain more code compared to reduction-focused methods, likely to ensure that functional correctness is not compromised. Such as in{\tt Util} benchmark, \textsc{Debop} only achieves 0.54 on the correctness score. 
Compared to ground truth (i.e., {\tt Cov-Perf}), which has similar correctness performance, we reduce 42.1\% and 15.6\% more elements on size and memory, respectively. This indicates that \toolname does not rely solely on coverage for debloating; it also refactors the program to create a more streamlined version. Additionally, \toolname's attack surface reduction (Atk) score is relatively low (0.20), which is similar to ground-truth {\tt Cov-Perf} (0.18). Similar results were observed on the {\tt SIR} dataset.
We believe this is due to the fact that the computation of the attack-surface is based on the number of return instructions, which in turn is highly correlated with the functionality of the program. It is difficult to reduce a large number of attack-surfaces while maintaining correctness.


Overall, \toolname stands out as a debloating approach that primarily focuses on preserving desired functionalities while reducing program size as much as possible. This aligns well with scenarios where correctness and robustness are essential, as it avoids aggressive debloating strategies that might compromise software functionality, thereby enhancing the practicability of \toolname.

\subsection{RQ2: Security of Debloated Programs}
\label{sec:security}
\begin{table}[!t]
  \centering
  \caption{Distribution of severity of vulnerabilities introduced by debloating tools.}
  \resizebox{0.98\linewidth}{!}{
    \begin{tabular}{ccccccccc}
    \toprule
    \multirow{2}[1]{*}{Tool} & \multicolumn{4}{c}{Reduced ($\uparrow$)}   & \multicolumn{4}{c}{Introduced ($\downarrow$)} \\
\cmidrule(lr){2-5}   \cmidrule(lr){6-9}       & Low   & Medium & High  & Sum   & Low   & Medium & High  & Sum \\
\midrule
    \textsc{Chisel} & 3.1   & 326.7  & 6.3   & 336.1  & 14.1  & 282.0  & 16.8  & 312.9  \\
    \textsc{Debop} & 5.1   & 351.4  & 14.9  & 371.3  & 42.7  & 366.5  & 38.9  & 448.1  \\
    Cov   & 4.9   & 348.4  & 4.9   & 358.2  & 41.2  & 369.5  & 34.2  & 444.8  \\
    CovF  & 14.2  & 346.8  & 14.2  & 375.2  & 33.0  & 357.0  & 38.6  & 428.6  \\
    CovA  & 4.2   & 339.5  & 11.2  & 354.8  & 20.7  & 327.3  & 16.9  & 364.9  \\
    Cov-Perf & 3.9 & 341.7 & 4.2 & 349.8 & 22.4 & 342.6 & 20.5 & 385.5 \\
    \midrule
    \toolname & \bf{16.3} & \bf{363.6} & \bf{16.5} & \bf{396.4} & \bf{5.7} & \bf{58.8} & \bf{11.6} & \bf{76.1} \\
    \bottomrule
    \end{tabular}%
    }
  \label{tab:security}%
\end{table}%

In this research question, we examine the number of vulnerabilities introduced by tools during the debloating process. As \textsc{Razor} performs debloating at the binary level, CodeChecker cannot directly analyze binary code. Therefore, we exclude \textsc{Razor} from this research question.

Table~\ref{tab:security} presents a comparison of various debloating tools in terms of the vulnerabilities they reduce and introduce across two benchmarks. We combine the results from both benchmarks to provide a larger benchmark, which offers a more robust comparison and reduces potential bias. The performance is assessed using two metrics: the number of vulnerabilities reduced (higher is better) and the number of vulnerabilities introduced (lower is better). The tool \toolname demonstrates a clear advantage over the other tools, achieving the highest number of reduced vulnerabilities across all severity levels, with a total of 396.4, including a particularly notable reduction in medium severity vulnerabilities (363.6). For high-severity vulnerabilities, compared to Cov, which removes the uncovered code snippets directly, \toolname removes 2.4 times more high-risk vulnerabilities. This indicates that the security advisor and LLM-based decision-maker can efficiently enhance the debloated program's security. 
Additionally, \toolname introduces significantly fewer vulnerabilities compared to the other tools, with a total of just 76.1. This represents a marked improvement over tools like \textsc{Debop} (448.1), Cov (444.8), and CovF (428.6). This demonstrates that \toolname is highly effective at minimizing the introduction of new vulnerabilities during the debloating process. Overall, these results highlight the effectiveness of \toolname in debloating programs while maintaining security, making it the most efficient and reliable tool in this comparison.

\subsection{RQ3: Ablation study}
We have demonstrated that a well-designed \toolname achieves state-of-the-art performance compared to existing studies. To understand the contribution of each component further, we created three variants of \toolname, each with one critical component removed, and evaluated their performance using the related splitting strategy with correctness-based metrics. We investigated the contribution of test augmentation (TA), security advisor (SA), and decision-maker (DM). Note that the debloating advisor is the foundation of our approach, hence we cannot remove it as without it other components cannot work properly. Specifically, for the variant without TA, we did not augment the original tests, meaning the \toolname would only observe a few tests and rely on them to debloat the program. For the variant without SA, we do not utilize the static analysis tool to analyze the debloated program, which means the decision-maker can only debloat the program relying on the info from the static analysis-based debloat advisor. For the variant without DM, we remove all potential unnecessary code marked by the debloat advisor and output it directly. Note that because invoking LLMs is time-consuming, we perform our ablation study on the {\tt Util} benchmark.
\begin{table}[!t]
  \centering
  \caption{Effectiveness of variants of \toolname on our evaluation datasets.}
  \resizebox{0.98\linewidth}{!}{
    \begin{tabular}{cccccccc}
    \toprule
    \multirow{2}[1]{*}{Variant} & \multirow{2}[1]{*}{Cor} & \multirow{2}[1]{*}{Rob} & \multicolumn{3}{c}{Reduction} & \multicolumn{2}{c}{Security} \\
\cmidrule(lr){4-6} \cmidrule(lr){7-8}   &    &  & Size  &  Mem & Atk & Reduced & Introduced \\
    \midrule
    -TA   &  0.63 & 0.64 & 0.76 & 0.68  & 0.28  & 192.5  & 38.1 \\
    -SA   &  0.95 & 0.94 & 0.55 & 0.54 &  0.20 &  157.8 & 147.6 \\
    -DM   &  0.97 & 0.95 & 0.40 & 0.45 & 0.18  &  162.4 & 172.5 \\
\toolname &  0.96 & 0.93 & 0.54  &  0.52 &  0.20 & 167.7 & 35.9 \\
    \bottomrule
    \end{tabular}%
    }
  \label{tab:ablation}%
\end{table}%

Table~\ref{tab:ablation} presents the results of an ablation study conducted to evaluate the effectiveness of different variants of \toolname on various performance metrics, including correctness, robustness, reduction, and security.
Specifically, the removal of the TA component significantly reduces correctness and robustness performance from 0.96 and 0.93 to 0.63 and 0.64, respectively. This decline suggests that the generated tests effectively cover most statements of the required features. However, with fewer tests in place, more code snippets are removed. While this increases code reduction, it also eliminates additional vulnerabilities.

For the variant without the SA component, correctness and robustness remain similar and are slightly higher than those of \toolname. This outcome may stem from security advisor suggestions that add security checks or prevent the deletion of security-related statements. The most significant impact appears in the number of introduced vulnerabilities, which rises from 35.9 to 147.6. Without the security advisor, the debloating process introduces far more vulnerabilities.

In the variant without the DM, uncovered statements are removed directly, affecting both the reduction and security metrics. The reductions in size, memory usage, and attack surface are less substantial than with \toolname, likely because the LLM-based decision maker refactors the program to optimize statement size and memory usage. The number of introduced vulnerabilities increases from 35.9 to 172.5, surpassing the vulnerabilities introduced by the variant without the SA. This indicates that the decision-maker can help prevent the introduction of vulnerabilities to some extent.

Overall, each component performs its intended function. The TA significantly enhances both correctness and robustness metrics. The SA helps \toolname mitigate the majority of vulnerabilities during the debloating process. The DM not only aids in reducing size and memory usage but also plays a role in preventing the introduction of new vulnerabilities.

\section{Discussion}
\subsection{Case Study}
\begin{figure}
    \centering
    \begin{subfigure}[b]{0.45\textwidth}
        \begin{lstlisting}[language=C, breaklines=true, basicstyle=\footnotesize]
_Bool c_isspace(int c) {
  return (c == 32 || c == 9 || c == 10 || c == 11 || c == 12 || c == 13);
}
        \end{lstlisting}
        \vspace{-4mm}
        \caption{Function generated by \toolname.}
        \label{subfig:case_study_isspace_full}
    \end{subfigure}
    \begin{subfigure}[b]{0.45\textwidth}
        \begin{lstlisting}[language=C]
_Bool c_isspace(int c) {
  return (c == 32 || c == 10) ? 1 : 0;
}
        \end{lstlisting}
        \vspace{-2mm}
        \caption{Function generated by \toolname without test augmentation.}
        \label{subfig:case_study_isspace_part}
    \end{subfigure}

    \caption{The debloated {\tt c\_isspace} function}
    \label{fig:case_study_isspace}
\end{figure}
Figure~\ref{fig:case_study_isspace} illustrates the {\tt c\_isspace} function generated by \toolname and its variant, highlighting \toolname's ability to debloat programs while preserving functionality and security. In particular, Figure~\ref{subfig:case_study_isspace_full} shows the function produced by \toolname. Compared to the version generated by input-based techniques (Figure~\ref{fig:moivation_vul}), \toolname maintains functionality through its documentation-guided test augmentation module, which increases the coverage of desired features. For example, the branch {\tt c==9}, removed by input-based techniques, is retained due to the augmented tests.
To further investigate the role of the decision-maker in maintaining security, we created a variant without test augmentation. In this variant, the absence of augmented tests left certain branches uncovered, resulting in a security vulnerability during the removal of uncovered code. As shown in Figure~\ref{subfig:case_study_isspace_part}, when specific branches (e.g., {\tt c==9}) are not covered by tests, a security vulnerability is introduced by existing debloating approaches (as in Figure~\ref{fig:suggestions_case}). In contrast, \toolname mitigates the issue by eliminating the use of the {\tt tmp} variable, avoiding the initialization of insecure values. This demonstrates the decision-maker's effectiveness in maintaining security.
In addition to preserving functionality and security, \toolname also achieves significant reduction. For instance, the original {\tt c\_isspace} function contains 21 statements, while the version generated by \toolname reduces this to a single statement. The removal of the temporary variable {\tt tmp} also reduces memory usage.
Overall, this case study demonstrates \toolname's capability to debloat programs effectively while maintaining functionality, ensuring security, and optimizing resource usage.




\subsection{The Impact of Different Static Analyzers}
In this study, we use CodeChecker as our security advisor to monitor security during debloating, due to its impressive performance in detecting vulnerabilities~\cite{charoenwet2024empirical}. We recognize that different static analysis tools may produce varying results because of their distinct methodologies. However, we believe the impact of using different static analyzers is limited for two main reasons. 
First, CodeChecker is an integrated tool that combines the results of several analyzers (e.g., Cppcheck~\cite{Cppcheck2024} and Infer~\cite{Infer2024}). This integration allows CodeChecker to cover a wider range of vulnerabilities compared to a single analyzer, thus reducing the potential bias that could arise from relying on just one tool. 
Second, in our evaluation, the security of all the investigated tools is assessed using the same static analysis tool. This ensures that the comparison remains fair and comparable.

\subsection{The Impact of Different Hyperparameters}
\label{subsec:hyperpara}
In this study, we set the temperature parameter $t$ of the invoked LLMs to 0.5 and limited the maximum number of iterations to 3 per function. To evaluate the impact of temperature on performance, we conducted additional experiments with different temperature settings: 0, 0.25, 0.5, 0.75, and 1. Higher temperatures encourage LLMs to generate more diverse outputs. Due to the time-intensive nature of invoking LLMs, these experiments were conducted on the {\tt Util} benchmark.

\begin{table}[!t]
  \centering
  \caption{The impact of different temperatures.}
  \resizebox{0.98\linewidth}{!}{
    \begin{tabular}{cccccccc}
    \toprule
    \multirow{2}[1]{*}{Temp.} & \multirow{2}[1]{*}{Cor} & \multirow{2}[1]{*}{Rob} & \multicolumn{3}{c}{Reduction} & \multicolumn{2}{c}{Security} \\
\cmidrule(lr){4-6} \cmidrule(lr){7-8}   &    &  & Size  &  Mem & Atk & Reduced & Introduced \\
    \midrule
    0     & 0.90 & 0.91 & 0.52 & 0.49 & 0.17 & 163.6 & 37.6 \\
    0.25  & 0.94 & 0.95 & 0.54 & 0.52 & 0.19 & 164.3 & 34.5\\
    0.5   & 0.96 & 0.94 & 0.54 & 0.52 & 0.20 & 167.7 & 35.9 \\
    0.75  & 0.95 & 0.95 & 0.55 & 0.53 & 0.21 & 165.5 & 36.6 \\
    1     & 0.95 & 0.94 & 0.54 & 0.53 & 0.20 & 168.4 & 37.2 \\
    \bottomrule
    \end{tabular}%
    }
  \label{tab:dis_temp}%
\end{table}%

Table~\ref{tab:dis_temp} summarizes the performance of \toolname under different temperature settings. Overall, the performance across most temperature values was consistent, with one notable exception at $t=0$, where the average output size was worse than other settings. We hypothesize that this is due to the low temperature limiting output diversity. For higher temperatures, we observed a slight decrease in correctness and a marginal increase in reduction. In the security dimension, performance varied without a clear trend related to temperature.
Considering the correctness and robustness, which we prioritize over reduction, we selected a temperature setting of $t=0.5$ for \toolname.

To determine the optimal maximum iterations per function, we analyzed the number of iterations required to successfully debloat functions in the {\tt Util} benchmark. Results show that 94.9\% of functions were successfully debloated within three iterations. This distribution indicates that setting the maximum number of iterations to three strikes a balance, ensuring the debloating of the majority of functions while minimizing excessive computational overhead.
\subsection{The Impact of Different LLMs}
\begin{table}[!t]
  \centering
  \caption{Effectiveness of \toolname with different LLM backends on {\tt Util} datasets.}
  \resizebox{0.98\linewidth}{!}{
    \begin{tabular}{cccccccc}
    \toprule
    \multirow{2}[1]{*}{LLM} & \multirow{2}[1]{*}{Cor} & \multirow{2}[1]{*}{Rob} & \multicolumn{3}{c}{Reduction} & \multicolumn{2}{c}{Security} \\
\cmidrule(lr){4-6} \cmidrule(lr){7-8}   &    &  & Size  &  Mem & Atk & Reduced & Introduced \\
    \midrule
    CodeLlama-13B & 0.89 & 0.91 & 0.48 & 0.46 & 0.16  & 160.4 & 42.3 \\
    GPT-4o & 0.95 & 0.93 & 0.52 & 0.52 & 0.21 &  175.6 & 29.2 \\
DeepSeek-V2.5 & 0.96 & 0.93 & 0.54 &  0.52 &  0.20 & 167.7 & 35.9 \\
    \bottomrule
    \end{tabular}%
    }
  \label{tab:different_llms}%
\end{table}%

To evaluate \toolname's performance across different LLM backends and mitigate the risk of potential data leakage from relying on a single LLM, we conduct an extensive experiment utilizing various LLMs. Due to time and token consumption constraints, this experiment is performed only on the {\tt Util} dataset. We examine two LLMs of different types (code-specific vs. general-purpose) and model sizes: CodeLlama-13B~\cite{roziere2023code} and GPT-4o~\cite{hurst2024gpt}. The evaluation results are presented in Table~\ref{tab:different_llms}. 
Specifically, when using CodeLlama-13B as the backend, correctness and robustness decrease while the number of introduced vulnerabilities increases. We attribute this to the fact that smaller models are less likely to recognize vulnerabilities in generated code than larger models (\eg GPT-4o and DeepSeek-V2.5), as observed in~\cite{bhatt2023purple}.  
In contrast, when utilizing GPT-4o as the backend, \toolname exhibits similar performance to the DeepSeek-V2.5 backend. Notably, GPT-4o achieves greater attack surface reduction and introduces fewer vulnerabilities during debloating. This aligns with findings that the GPT series of models undergo specialized security enhancements~\cite{openai2024safety} and have been shown to generate more secure code in prior studies~\cite{bhatt2023purple,sajadi2025llms}.

\subsection{The Efficiency of Approaches}
\begin{table}[!t]
  \centering
  \caption{The Efficiency of investigated approaches}
  \resizebox{1\linewidth}{!}{
      \begin{tabular}{cccccccc}
    \toprule
    Tool  & \textsc{Chisel} & \textsc{Debop} & \textsc{Razor} & {Cov} & {CovF} & {CovA} & {\toolname} \\
    Avg.(min) & 284.3 & 324.8 & 0.2   & 0.4   & 3.3   & 0.6   & 84.7 \\
    \bottomrule
    \end{tabular}%
  }

  \label{tab:efficiency}%
\end{table}%

To assess the efficiency, we calculate the time consumption of the debloating process, defined as the duration required to generate a debloated program across the {\tt Util} and {\tt SIR} dataset. Table~\ref{tab:efficiency} reveals that \textsc{Debop} and \textsc{Chisel} demonstrate the highest computational overhead, averaging 324.8 and 284.3 minutes respectively. This aligns with their use of stochastic optimization and reinforcement learning, both being resource-intensive methodologies. \textsc{Razor}'s heuristic-driven approach based on execution path analysis proves markedly more efficient, completing debloating in 0.2 minutes on average. \toolname achieves an optimal balance between efficacy and efficiency at 84.7 minutes per program. Notably, these results reflect single-threaded execution; parallel implementation could substantially reduce \toolname's debloating time through concurrent processing capabilities.

\subsection{Difference Between Test Augmentation and Fuzzing}
In this paper, we introduce a documentation-guided test augmentation component designed to generate test cases. While fuzzing techniques are also employed for test generation, there are significant differences between test augmentation and fuzzing that make existing fuzzing methods unsuitable for augmenting tests. Specifically, test augmentation in \toolname aims to maximize coverage of user-specified features derived from documentation, ensuring critical functionalities are preserved. In contrast, fuzzers prioritize exploring edge cases to uncover bugs, often through mutation of existing tests. While both generate tests, their objectives differ fundamentally: LEADER’s component explicitly guides LLMs to generate tests aligned with desired features, whereas fuzzing focuses on code coverage breadth. This distinction ensures \toolname's tests are feature-preserving rather than exploring edge cases.

\subsection{Threats to Validity}

\subsubsection{Internal Validity} 
One potential internal threat stems from potential data leakage. This raises questions about how LLMs handle code debloating: do they identify and remove unreachable code, or do they simply memorize the debloated program and output it directly?
However, this task is inherently less affected by data leakage for two reasons. First, the existing programs in the training set cannot explicitly benefit the program debloating task, which distinguishes it from other code-related tasks such as program repair or code generation. In program repair, for instance, it is possible for LLMs to output an existing fixed program rather than repair the bug. However, program debloating is complex, and there is no ready-made debloated program that can be leveraged directly. 
Second, although BusyBox~\cite{busybox_website} provides simplified versions of several common Unix tools for embedded systems, these tools are usually redesigned rather than simplified from their Unix versions. This means LLMs cannot directly leverage the source code from BusyBox. For example, we investigated the implementation of {\tt Sort} in both BusyBox and Unix. We found no simplified functions in BusyBox, as no functions share the same name in these two versions. This further confirms that LLMs cannot leverage the source code from BusyBox. Thus, we are confident that data leakage is not a key factor in our conclusion.

Another potential threat to internal validity arises from the evaluation of correctness and robustness. For correctness evaluation, we split the test suite into two subsets: one for debloating and one for testing. However, the limited number of tests may be insufficient for a comprehensive evaluation. A potential solution is to employ the test augmentation component to generate additional tests. While generated tests may include low-quality cases, such as redundant or invalid tests, they have minimal impact on the debloating process since it focuses on code coverage across all provided tests. Nonetheless, such low-quality tests could introduce bias during evaluation. For instance, multiple generated tests may assess the same function or feature, leading to an overemphasis on certain code paths. This could cause a debloated program to score artificially high if it passes redundant tests, even if it fails to maintain other critical functionalities. To mitigate this bias, we use a manually curated test set that ensures a balanced representation of test cases for each feature. This approach guarantees fairness across the tools being compared, as all are evaluated against the same criteria.
For robustness evaluation, we employ the AFL~\cite{AFL} fuzzing tool, setting a five-hour execution limit in alignment with previous studies~\cite{xin2022studying,heo2018effective}. Empirical findings~\cite{fioraldi2023dissecting} indicate that AFL attains over 95\% code coverage within the initial five hours of a 23-hour execution. Thus, we consider this time limit is sufficient for a fair and meaningful evaluation.

\subsubsection{External Validity}
One threat to external validity is the generality of \toolname across programming languages.
Following previous studies~\cite{xin2022studying, heo2018effective, xin2020program,qian2019razor}, we evaluate the approaches on two C language benchmarks.
Although \toolname is designed to be language-agnostic and could be implemented for other languages, its effectiveness on other languages remains unknown.

\section{Related Work}
\label{sec:relatedWork}
{\bf Program Reduction}. Program reduction is a closely related research domain to program debloating and is widely used for testing and debugging. These techniques~\cite{zhang2024lpr,xu2023pushing,wang2021probabilistic,tian2023caching} aim to facilitate compiler debugging by minimizing bug-triggering programs. Formally, given a program $P$ that exhibits a specific property $\psi$ (e.g., a C program that crashes GCC during compilation), the objective of program reduction is to minimize $P$ into a smaller variant $P'$ that still preserves property $\psi$.  

Compared to program debloating, program reduction has a more precisely defined specification. It focuses exclusively on preserving bug-triggering behaviors, whereas program debloating must infer implicit specifications beyond those encoded in test cases. Consequently, program debloating requires a specialized design to infer specifications from the provided tests. As a result, existing program reduction techniques cannot be directly applied to debloating.

{\bf Program Repair.} Similar to program debloating techniques, program repair approaches also rely on a set of tests to identify a program that satisfies the specifications defined by these tests. Based on the methodology used to discover the desired program, these approaches can be broadly categorized into three main types: search-based~\cite{qi2014strength,jiang2018shaping,liu2018mining,weimer2009automatically}, constraint-based~\cite{nguyen2013semfix,mechtaev2016angelix,xuan2017nopol,long2015staged}, and learning-based~\cite{ye2022neural,lutellier2020coconut,xia2022less,lin2024one}.

\section{Conclusion}
\label{sec:conc}
This paper presents \toolname, a program debloating framework leveraging LLMs to maintain functionality and security during debloating. By combining a documentation-guided test augmentation module with a multi-advisor-aided debloating module, \toolname addresses key limitations of existing techniques. Extensive evaluations show \toolname achieves a 95.5\% test case pass rate, reduces program size by 42.5\%, and significantly decreases both introduced and pre-existing vulnerabilities compared to CovA. These results highlight \toolname's potential to set a new standard for secure and effective program debloating.




%


\bibliographystyle{plain} 
\bibliography{bib/references}

\end{document}